\documentstyle[preprint,aps,floats,epsf]{revtex}
\begin{document}


\preprint{$
\begin{array}{l}
\mbox{UB-HET-97-01}\\[-3mm]
\mbox{UCD--97--02}\\[-3mm]
\mbox{MADPH-97-986}\\[-3mm]
\mbox{February~1997} \\   [1cm]
\end{array}
$}

\title{$W\gamma\gamma$ Production at the Fermilab Tevatron Collider: 
Gauge Invariance and Radiation Amplitude Zero\\[1.cm]}

\author{U. Baur$^a$, T. Han$^b$, N.~Kauer$^c$, R. Sobey$^b$, 
and D.~Zeppenfeld$^c$\\[0.5cm]}

\address{$^a$Department of Physics,
State University of New York, Buffalo, NY  14260, USA\\
$^b$Department of Physics,
University of California, Davis, CA 95616, USA\\
$^c$Department of Physics,
University of  Wisconsin,  Madison, WI 53706, USA\\[1.5cm]}

\maketitle
\begin{abstract}
The electroweak process $p\bar p\to \ell^\pm\nu\gamma\gamma$ is
calculated at tree level, including finite
$W$ width effects. In order to obtain a gauge invariant amplitude, the
imaginary parts of $WW\gamma$ triangle graphs and
$WW\gamma\gamma$ box diagrams have to be included, in addition to
resumming the imaginary contributions to the $W$ vacuum polarization. 
We demonstrate the existence of a radiation amplitude zero
in $p\bar p\rightarrow W^\pm \gamma\gamma\to \ell^\pm\nu\gamma\gamma$, 
and discuss how it may be observed in correlations of the
$\gamma\gamma$ and lepton rapidities at the Fermilab Tevatron. 
\end{abstract}


\newpage

\narrowtext

\section{Introduction}

Radiative $W$ production and decay at hadron colliders is an important
testing ground for the Standard Model (SM). The simplest process, $q\bar
q'\to\ell^\pm\nu\gamma$, allows the measurement of the $WW\gamma$ three 
gauge boson coupling at large photon transverse 
momenta~\cite{BROWN,CHH,BZ,WGEXP}. In addition, this process is of special
interest due to the presence of a zero in the amplitude of the parton
level process $q\bar q'\to W\gamma$ \cite{BROWN,RAZ,theorem}.
At small transverse momenta of the photon
or when the photon is emitted collinearly to the final state charged
lepton, this process needs to be fully understood when trying to extract
a precise value of the $W$ boson mass from Tevatron data. Approximately
24\% (13\%) of all $W\to e\nu$ ($W\to\mu\nu$) events contain a photon
with a transverse momentum ($p_T^\gamma$) larger than 100~MeV~\cite{BK,bzprl}, 
the approximate threshold of the electromagnetic calorimeter of the CDF 
and D\O\ detectors. Radiative $W$ decay events shift the $W$ mass
by about 65~MeV in the electron, and by approximately 170~MeV in the
muon channel~\cite{CDFWmass,D0Wmass}. 

For similar reasons, the process $q\bar q'\to\ell^\pm\nu\gamma\gamma$ is
interesting. At large photon transverse momenta, $W\gamma\gamma$
production is sensitive to the structure of the $WW\gamma\gamma$
quartic coupling~\cite{arg}. Furthermore, as a consequence of a general
theorem~\cite{theorem} one expects a radiation zero in the SM $q\bar q'\to
W\gamma\gamma$ amplitude. Two photon radiation is expected to have a
non-negligible effect on the $W$ mass extracted from future high precision 
Tevatron data because approximately 0.8\% of all $W\to\mu\nu$ events
are expected to contain two well separated photons with $p_T^\gamma
>100$~MeV~\cite{BHSZ}.  Finally, $W\gamma\gamma$ production is an
irreducible background to associated production of a $W$ and a Higgs
boson in hadronic collisions, if the Higgs boson decays into two 
photons~\cite{james}.

In $p\bar p$ collisions at a center of mass energy of 2~TeV, the total 
cross section for $W^\pm\gamma\gamma$ production is approximately 
4.6~fb, when only considering leptonic decays, $W\to\ell\nu$ ($\ell=e,\,\mu$),
and $p_T^\gamma>10$~GeV and $|\eta_\gamma|<2.5$ ($\eta$ being the 
pseudorapidity)~\cite{hansobey}. Upgrades of the Tevatron accelerator 
complex (TeV33), beyond the Main Injector project, could yield an 
overall integrated luminosity of ${\cal O}(30$~fb$^{-1}$)~\cite{marriner}, 
making a study of $W\gamma\gamma$ production a
realistic goal in the TeV33 era. The hadronic decay modes of the $W$ 
will be difficult to
observe due to the QCD $jj\gamma\gamma$ background~\cite{BHOZ}. We
therefore concentrate on the leptonic decays of the $W$ boson, and
calculate the helicity amplitudes for the complete process
\begin{equation}
q\bar q'\to\ell^\pm\nu\gamma\gamma,
\end{equation}
including Feynman diagrams where one or both photons are emitted from
the final state charged lepton line. In a realistic simulation, these 
diagrams, together with finite $W$ width effects, need to be taken into
account. 

When including finite $W$ width effects, some care is needed to preserve
gauge invariance. Replacing the $W$ propagator, $1/(q^2-m_W^2)$, by a 
Breit-Wigner form, $1/(q^2-m_W^2+im_W\Gamma_W)$, will disturb the gauge 
cancellations between the individual Feynman graphs and thus lead to an 
amplitude which is not electromagnetically gauge invariant. In
addition, a constant imaginary part in the 
inverse propagator is ad hoc: it results from fermion loop contributions 
to the $W$ vacuum polarization and the imaginary part should vanish for 
space-like momentum transfers. In Ref.~\cite{bzprl} it was demonstrated 
how this problem is solved for $W\gamma$ production by including the 
imaginary part of $WW\gamma$ vertex corrections in addition to the 
resummation of the $W$ vacuum polarization contributions. Here, we
generalize the result of Ref.~\cite{bzprl} to $W\gamma\gamma$ production, 
and show that a gauge invariant amplitude for the process 
$q\bar q'\to\ell^\pm\nu\gamma\gamma$ is obtained by also including 
the imaginary part of the $WW\gamma\gamma$ box corrections.
Extending the argument, a gauge
invariant amplitude for $q\bar q'\to W+n\gamma$, $n>2$, can be obtained
by implementing the corrected $WW\gamma$ and $WW\gamma\gamma$ vertices
together with the resummed $W$ vacuum polarization contributions.
No higher $WWn\gamma$ vertex functions need to be considered.
Our analysis of gauge invariance for $W\gamma\gamma$ production is 
described in Sec.~II.

The existence of a radiation zero in the process 
$q\bar q'\to W\gamma$ has been well known for more than fifteen 
years~\cite{RAZ,theorem}. All SM helicity amplitudes for the process
$q\bar q' \to W\gamma$ vanish at
\begin{eqnarray}
\label{eq:raz}
\cos \theta^*_W = \cos \theta_{0W}^* = \frac {Q_q + Q_{q'}} {Q_q-Q_{q'}},
\end{eqnarray}
where $\theta^*_W$ is the angle between the $W$ and the incoming quark $q$,
in the parton center of mass frame. A theorem~\cite{theorem} then
predicts that the process $q \bar q' \to W+ n\gamma$, $n>1$, exhibits a 
radiation zero for the same scattering angle $\cos \theta_{0W}^*$, if 
the $n$ photons are collinear. In 
Sec.~III, we numerically demonstrate the existence of this radiation zero
in $W\gamma\gamma$ production.

In practice radiation zeros in hadronic collisions are
difficult to observe. In the $W\gamma$ case, the ambiguity 
in reconstructing the parton center of mass frame and in identifying
the quark momentum direction represents a major complication 
in the extraction of the $\cos\theta^*_W$ distribution~\cite{CHH}. 
Higher order QCD corrections~\cite{NLO,NLOTWO} and finite $W$ width 
effects, together with photon radiation from the final state lepton 
line, transform the zero to a dip~\cite{VALENZUELA}. Finite detector 
resolution effects further dilute the radiation zero. 
The twofold ambiguity in reconstructing the parton center of mass 
frame originates from the
nonobservation of the neutrino arising from $W$ decay. Identifying the
missing transverse energy with the transverse momentum of the
neutrino, the unobservable longitudinal neutrino 
momentum, $p_L(\nu)$, and thus the parton center of
mass frame, can be reconstructed by imposing the constraint that the
neutrino and charged lepton four momenta combine to form the $W$ rest
mass~\cite{STROUGHAIR}. The resulting quadratic equation, in general, has
two solutions. 
Finally, determining the $\cos\theta^*_W$ distribution requires
measurement of the missing transverse energy in the event. In future 
Tevatron runs, one expects up to ten
interactions per bunch crossing~\cite{marriner}. Multiple interactions
per crossing significantly worsen the missing transverse energy
resolution, and thus tend to wash out the dip caused by the radiation
zero. 

For $W\gamma\gamma$ production, the same problems arise. In addition, it 
is very difficult to experimentally separate two collinear photons, and, 
thus, to distinguish the $W\gamma\gamma$ signal from $W\gamma$ events and 
from the $W+$~jets background, where one of the jets fluctuates into 
a $\pi^0$ which decays into two almost collinear photons. One
therefore has to search for a signal of the radiation zero which
survives an explicit photon--photon separation requirement. 

In Ref.~\cite{rapraz} it was found that lepton--photon rapidity
correlations offer the best chance to observe the radiation zero in
$W\gamma$ production. The distribution of the rapidity difference $\Delta 
y(\gamma,\ell)=y_\gamma - y_\ell$ clearly displays the SM radiation
zero. It does not require knowledge of the longitudinal momentum of the 
neutrino, and so automatically avoids the problems described above.
In Sec.~III we show that the concept of
rapidity correlations as a tool to search for radiation zeros can be
generalized to the $W\gamma\gamma$ case. The $\Delta y(\gamma\gamma,\ell)$
distribution with $\cos\theta_{\gamma\gamma}>0$, where 
$\theta_{\gamma\gamma}$ is the opening angle between the two photons in 
the laboratory system, clearly displays the SM radiation zero even when 
one requires two well separated photons, provided that cuts are imposed 
which reduce the background from radiative $W$ decays. 

It is sometimes useful to compare distributions for $W\gamma\gamma$ and
$Z\gamma\gamma$ production. Simultaneously with the calculation of the
process $q\bar q'\to\ell^\pm\nu\gamma\gamma$, we therefore also present
results for $q\bar q\to\ell^+\ell^-\gamma\gamma$ production in Sec.~III.
Section~IV contains some concluding remarks.

\section{Finite Width Effects and Gauge Invariance in $W\gamma\gamma$
Production}

\begin{figure}
\begin{center}
\epsfxsize=5.25in
\epsfysize=5.25in
\epsfxsize=4.75in\hspace{0in}\epsffile{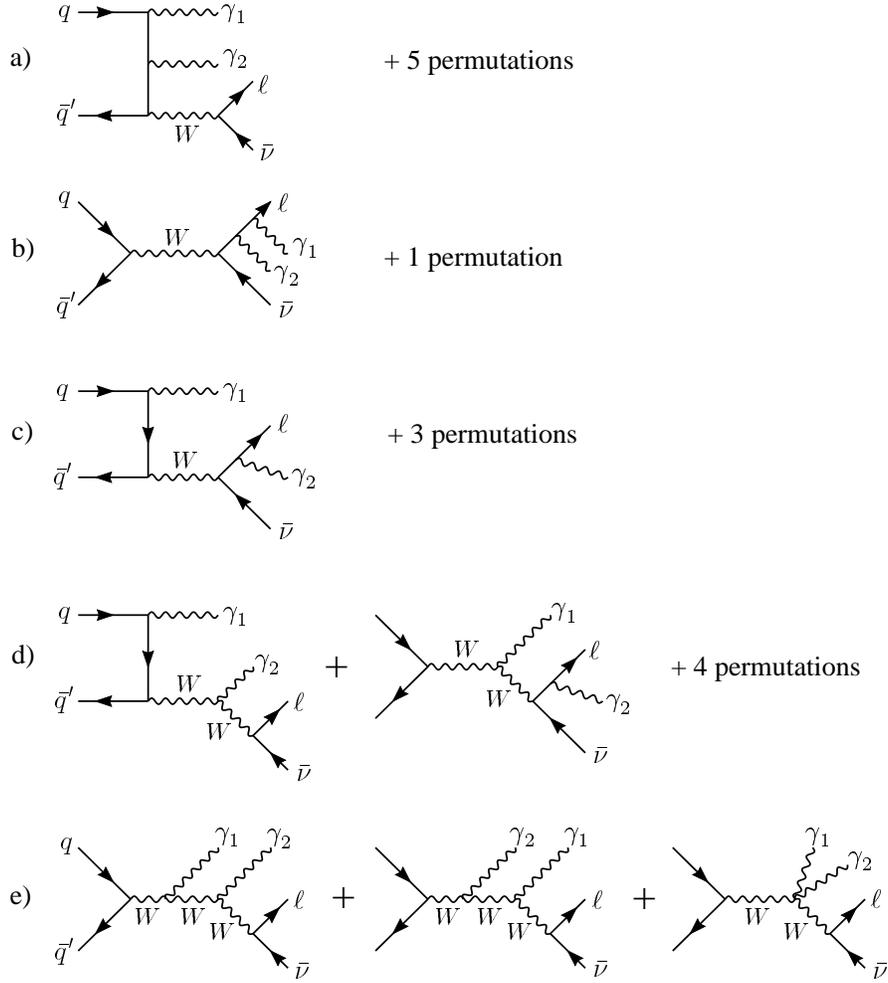}
\vspace*{0.5in}
\caption{
Feynman diagrams for the process $q\bar q' \to \ell\nu\gamma\gamma$. 
Permutations of the final state photons and the $W$ boson are not shown 
explicitly.}
\label{fig:feyn}
\end{center}
\end{figure}

At the parton level, the reaction $p\bar p\to\ell^\pm\nu\gamma\gamma$
proceeds via the Feynman diagrams shown in Fig.~\ref{fig:feyn}. Besides
the diagrams for $W\gamma\gamma$ production, graphs describing $W\gamma$
production followed by $W\to\ell\nu\gamma$ contribute, and also the 
radiative decay $W\to\ell\nu\gamma\gamma$. When finite $W$ width effects 
are included, the three reactions can no longer be distinguished, and 
the full set of Feynman diagrams must be taken into account. 
To calculate the helicity
amplitudes for the process $q\bar q'\to\ell^\pm\nu\gamma\gamma$ we have
used the framework of Refs.~\cite{hz} and~\cite{alan}. The result was
then compared numerically with the amplitudes obtained using the 
MADGRAPH/HELAS program~\cite{HELAS,MadGraph} which generates helicity 
amplitudes automatically. All quarks and leptons were assumed to be 
massless in our numerical calculations.

A naive implementation of finite $W$ width effects, by replacing the $W$
propagator by a Breit--Wigner form with momentum dependent width, leads
to a violation of electromagnetic gauge invariance~\cite{lopez}
and the resulting cross sections cannot be trusted. One encounters the 
same problem when gauge invariance is restored in an ad hoc 
manner~\cite{bzprl,vO}. Finite width effects are included in a tree level
calculation by resumming the imaginary part of the $W$ vacuum polarization.
Gauge boson loops ($W\gamma$ and $WZ$) only contribute above the $W$-mass
pole and are suppressed by threshold factors~\cite{doreen}. They can 
safely be neglected at the desired level of accuracy and only fermion 
loops need to be considered. Neglecting the fermion masses in the 
loops, the transverse part of the $W$ vacuum polarization receives an 
imaginary contribution
\begin{equation}\label{Pi_W^T}
Im\, \Pi_W^T(q^2) = \sum_f {g^2\over 48\pi} q^2
= q^2 {\Gamma_W\over m_W} = q^2 \gamma_W \; ,
\end{equation}
while the imaginary part of the longitudinal piece vanishes. In the
unitary gauge the $W$ propagator is thus given by
\begin{eqnarray}
D_W^{\mu\nu}(q) & = &{-i \over q^2-m_W^2 + i Im\, \Pi_W^T(q^2)}
\left( g^{\mu\nu}-{q^\mu q^\nu \over q^2}\right)
+{i\over m_W^2 - i Im\, \Pi_W^L(q^2)}\; {q^\mu q^\nu \over q^2}
\nonumber \\[3.mm]
& = & {-i \over q^2-m_W^2 + i q^2 \gamma_W}
\left( g^{\mu\nu}-{q^\mu q^\nu \over m_W^2}
(1+ i \gamma_W )   \right)\; . \label{Wprop}
\end{eqnarray}

\begin{figure}
\begin{center}
\epsfxsize=4.25in
\epsfysize=3.25in
\epsfxsize=4.25in\hspace{0in}\epsffile{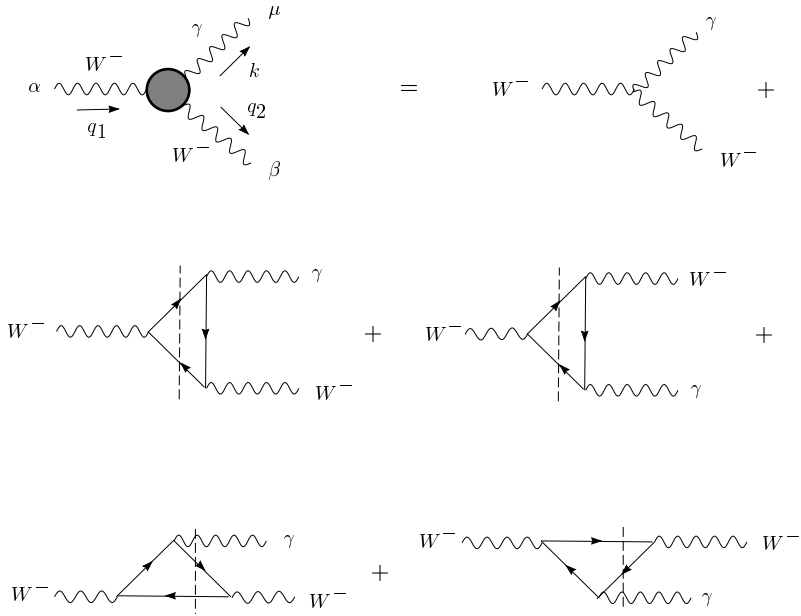}
\caption{
The effective $WW\gamma$ vertex as needed in 
the tree level calculation of $\ell\nu\gamma\gamma$ production.} 
\label{fig:feyntriangle}
\end{center}
\end{figure}

\begin{figure}
\begin{center}
\epsfxsize=4.75in
\epsfysize=4.0in
\epsfxsize=4.in\hspace{0in}\epsffile{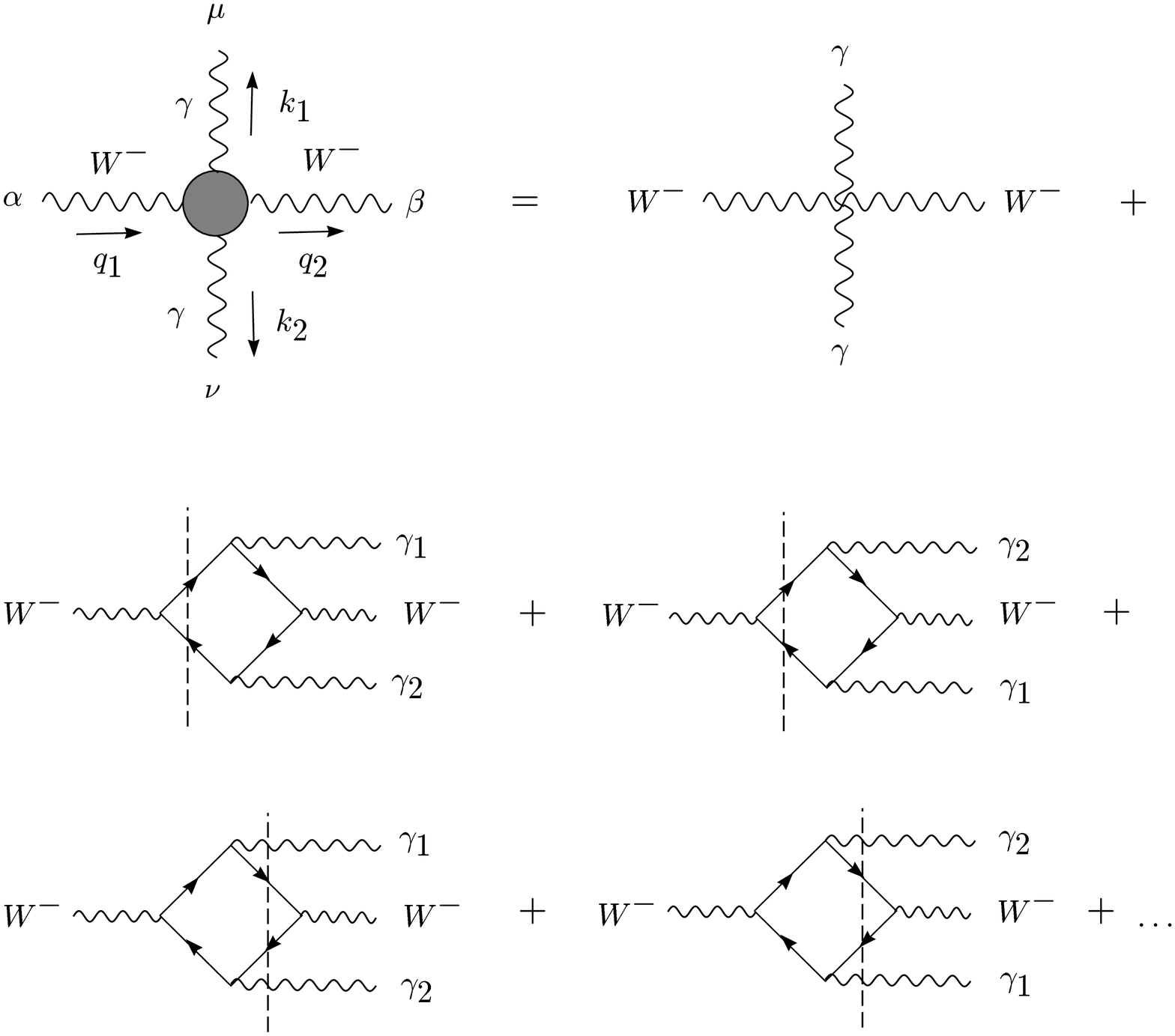}
\vspace*{0.2in}
\caption{
The Feynman graphs contributing to the effective $WW\gamma\gamma$ vertex of 
Eq.~(\protect\ref{WWggvertex}). Only four of the 24 cut box diagrams which
contribute to the imaginary part of the one-loop $WW\gamma\gamma$ vertex
are shown.}
\label{fig:feynbox}
\end{center}
\end{figure}

A gauge invariant expression for the amplitude of the process
$q\bar q' \to \ell^\pm\nu\gamma\gamma$ is obtained by attaching the final
state photons in all possible ways to all charged particle propagators
in the Feynman graphs. To be specific, we
shall concentrate on the $\ell^-\bar\nu\gamma\gamma$ final state in the
following. In addition to radiation off the external fermion 
lines and radiation off the $W$ propagators, the photons must be
attached to the charged fermions inside the $W$ vacuum polarization
loops, leading to the fermion triangle and box graphs of 
Figs.~\ref{fig:feyntriangle} and~\ref{fig:feynbox}. Since we are only 
keeping the imaginary part of $\Pi_W^T(q^2)$,
consistency requires including the imaginary parts of the
triangle and box graphs only. These imaginary parts are obtained by 
cutting the triangle and box graphs into on-shell intermediate states 
in all possible ways, as shown in the figures. 

For the momentum flow displayed in Figs.~\ref{fig:feyntriangle} 
and~\ref{fig:feynbox}, the tree level $WW\gamma$ and
$WW\gamma\gamma$ vertices are given by the familiar expressions
\begin{eqnarray}
-ie\Gamma_0^{\alpha\beta\mu} &=& -i e \left( (q_1+q_2)^\mu g^{\alpha\beta}
                            - (q_1+k)^\beta g^{\mu\alpha}
                            + (k-q_2)^\alpha g^{\mu\beta} \right)\; ,
\label{LOWWgvertex}
\\
-ie\Gamma_0^{\alpha\beta\mu\nu} &=& -i e^2 \left(2 g^{\alpha\beta} g^{\mu\nu} 
- g^{\alpha\mu} g^{\beta\nu} - g^{\alpha\nu} g^{\beta\mu} \right)\; .
\label{LOWWggvertex}
\end{eqnarray}
For the triangle graphs of Fig.~\ref{fig:feyntriangle} the momentum
configuration, $q_1^2>q_2^2,\;k^2=0$, is the same as the one encountered 
in the $q\bar q'\to\ell^\pm\nu\gamma$ case. Neglecting fermion masses, 
the full $WW\gamma$ vertex is then given by~\cite{bzprl}
\begin{equation}\label{WWgvertex}
\Gamma^{\alpha\beta\mu} =
\Gamma_0^{\alpha\beta\mu}\left(1+\sum_f {ig^2\over 48 \pi}\right) =
\Gamma_0^{\alpha\beta\mu}\left(1+i\,{\Gamma_W\over m_W}\right) =
\Gamma_0^{\alpha\beta\mu}\left(1+i\gamma_W \right)\; .
\label{EQ:TRI}
\end{equation}
Non-zero fermion masses, $m_f>0$, introduce corrections to 
Eq.~(\ref{EQ:TRI}) and generate 
axial vector contributions to the $WW\gamma$ vertex which are
proportional to $m_f^2/q_1^2$ and $m_f^2/q_2^2$~\cite{pest}. They can
be neglected at the desired level of accuracy for the lepton and the light 
quark doublets. Top-bottom loops do not contribute to the imaginary part
of the $WW\gamma$ vertex below threshold, {\it i.e.} 
for $q_i^2 < (m_t+m_b)^2$. In the imaginary part of the $WW\gamma\gamma$ 
vertex they are either absent for $q_i^2 < (m_t+m_b)^2$ or are suppressed 
by powers of the top quark mass. These massive loops are not needed for 
the restoration of electromagnetic gauge invariance
and can be neglected close to the $W$ pole.

An evaluation of the 24 cut box diagrams of Fig.~\ref{fig:feynbox} 
yields a result similar to that of Eq.~(\ref{EQ:TRI})~\cite{kauer}. 
For vanishing fermion
masses, each fermion doublet $f$, irrespective of its hypercharge, adds 
$i(g^2/48\pi)\Gamma_0^{\alpha\beta\mu\nu}$ to the tree level 
$WW\gamma\gamma$ vertex $\Gamma_0^{\alpha\beta\mu\nu}$. In the phase
space region $q_1^2>q_2^2,\;k_1^2=k_2^2=0$, the full $WW\gamma\gamma$ 
vertex is thus given by
\begin{equation}\label{WWggvertex}
\Gamma^{\alpha\beta\mu\nu} =
\Gamma_0^{\alpha\beta\mu\nu}\left(1+i\gamma_W \right)\; .
\end{equation}
In the expressions for the two vertices, terms proportional to $k_1^\mu$ 
or $k_2^\nu$ have been dropped. Such terms will be contracted with the photon
polarization vectors $\varepsilon^{*\mu}(k_1)$ or
$\varepsilon^{*\nu}(k_2)$, 
or a conserved electromagnetic current and hence vanish in the amplitude. 
Similarly, terms proportional to $q_1^\alpha$ are dropped since, in the 
massless quark limit, the $W$ couples to a conserved quark current. No such
assumption is made for the $W$-decay leptons, and hence our expressions are
valid when including finite charged lepton masses. For off-shell
photons or space-like $W$-bosons, more complicated expressions are 
obtained~\cite{argyres}.

By construction, the resulting amplitude for the process
$q\bar q' \to \ell^- \bar\nu  \gamma \gamma$ should be gauge invariant.
Indeed, gauge invariance of the full amplitude can be traced to the
electromagnetic Ward identities 
\begin{equation}\label{ward}
k_\mu {\Gamma_{\alpha\beta}}^\mu = \left(
iD_W\right)^{-1}_{\alpha\beta}(q_1)
- \left( iD_W\right)^{-1}_{\alpha\beta}(q_2)\; ,
\end{equation}
between the $WW\gamma$ vertex and the inverse $W$ propagator~\cite{lopez}
and
\begin{equation}\label{ward34}
k_1^{\mu} \Gamma_{\alpha\beta\mu\nu}(q_1,q_2,k_1,k_2) 
= e \left(\Gamma_{\alpha\beta\nu}(q_1,q_2+k_1,k_2) 
- \Gamma_{\alpha\beta\nu}(q_1-k_1,q_2,k_2) \right) \; ,
\end{equation}
relating three- and four-point functions. Since
\begin{equation}\label{ward1}
k_\mu \Gamma^{\alpha\beta\mu} = \left(
( q_1^2 g^{\alpha\beta} - q_1^\alpha q_1^\beta ) -
( q_2^2 g^{\alpha\beta} - q_2^\alpha q_2^\beta ) \right)
\left(1+i\gamma_W\right)\;,
\end{equation}
and
\begin{equation}\label{ward2}
\left( iD_W\right)^{-1}_{\alpha\beta}(q) =
\left( q^2-m_W^2+iq^2\gamma_W \right)
\left( g_{\alpha\beta} - {q_\alpha q_\beta \over q^2} \right)
-m_W^2 {q_\alpha q_\beta \over q^2}\; ,
\end{equation}
the Ward identity of Eq.~(\ref{ward}) is satisfied for 
the $W$ propagator and $WW\gamma$
vertex of Eqs.~(\ref{Wprop}) and~(\ref{WWgvertex}). Similarly the Ward 
identity for the $WW\gamma\gamma$ vertex is verified for the explicit
three- and four-point functions of Eqs.~(\ref{WWgvertex}) 
and~(\ref{WWggvertex}).

Extending this analysis to the $WW\gamma\gamma\gamma$ vertex, the 
relevant Ward identity relating $WW\gamma\gamma\gamma$- and 
$WW\gamma\gamma$ vertices is given by
\begin{equation}
k_{1\mu} \Gamma_{\alpha\beta\nu\rho}^\mu(q_1,q_2,k_1,k_2,k_3) 
= e \left(\Gamma_{\alpha\beta\nu\rho}(q_1,q_2+k_1,k_2,k_3) 
- \Gamma_{\alpha\beta\nu\rho}(q_1-k_1,q_2,k_2,k_3) \right) \; .
\end{equation}
The right hand side vanishes for the tree-level $WW\gamma\gamma$
vertex and thus also for the fermion-one-loop corrected vertex of
Eq.~(\ref{WWggvertex}). This means that the amplitude for the three
photon process $q\bar q' \to \ell^\pm \nu \gamma \gamma \gamma$ is rendered
gauge invariant by implementing the corrected $WW\gamma$
and $WW\gamma\gamma$ vertex functions only, but without taking into account
any $WW\gamma \gamma \gamma$ one-loop correction. The argument can
immediately be generalized to an arbitrary number of final state photons.
For hard, non-collinear photon emission this is mostly of theoretical
interest, however, since the cross section for $W\gamma \gamma \gamma$
production, already, is expected to be too small to be observed
even at a high luminosity Tevatron.

Returning to the calculation of the $q\bar q'\to\ell\nu\gamma\gamma$
amplitude, a gauge invariant result is obtained
by replacing all $W$-propagators, $WW\gamma$ and $WW\gamma\gamma$ vertices
in the Feynman graphs of Fig.~\ref{fig:feyn} by the full expressions of 
Eqs.~(\ref{Wprop}), (\ref{WWgvertex}) and (\ref{WWggvertex}), respectively.
Formally, these expressions include the imaginary parts of up to two loops 
in the vertices of Fig.~\ref{fig:feyn}(e). However, the Dyson resummation 
of the $W$-propagators already constitutes a mixing of all orders of 
perturbation theory and thus the appearance of several vertex loops should
be no surprise. This result is obtained naturally by attaching the two 
photons in all possible ways to either one of the fermion
loops or to one of the lowest order $W$ propagators in the zero, one, 
two etc. fermion bubble graphs contributing to the Dyson resummed process
$q\bar q' \to \ell\nu$: the remaining sum over $W$ vacuum polarization graphs
restores the full $W$ propagator of Eq.~(\ref{Wprop}) on either side of a 
triangle graph, a $WW\gamma$ vertex, the box graph, or the $WW\gamma\gamma$ 
vertex. After resummation, one therefore obtains the Feynman graphs of 
Fig.~\ref{fig:feyn} where any $WW\gamma(\gamma)$
vertex is given by the sum of the lowest order vertex and the imaginary
part of the triangle (box) graphs, as defined in 
Figs.~\ref{fig:feyntriangle} and~\ref{fig:feynbox}. 

Finally note that conservation of the final state lepton 
current has not been assumed anywhere, {\it i.e.} terms proportional 
to $q_2^\beta$ have been kept throughout. 
Thus our calculation is correct for massive final state
leptons and the emission of two photons collinear with the charged
final state lepton can be simulated with the resulting code~\cite{BHSZ}.
Alternative approaches in treating unstable gauge bosons in a gauge 
invariant way have been discussed in Ref.~\cite{stuart}.

\section{Searching for the Radiation Zero in $W\gamma\gamma$ Production
at the Tevatron}

\subsection{Input Parameters and Detector Simulation}

We now study in detail the radiation zero in $q\bar q'\to
W\gamma\gamma$ predicted by the SM, for $p\bar p$ collisions at
$\sqrt{s}=2$~TeV. To simplify the discussion, we shall concentrate on
the $W^-\gamma\gamma$, $W^-\to e^-\bar\nu$ channel. In $p\bar p$
collisions, the total cross sections for $W^-\gamma\gamma$ and
$W^+\gamma\gamma$ production are equal. Angular and rapidity
distributions for the $W^+$ case can be obtained by a sign change of the
variable. The parameters used in our numerical
simulations are $m_{W}=80.22$~GeV, $m_{Z}=91.187$~GeV, 
and $\alpha_{em}=1/128$. We use the parton distribution functions set A of
Martin-Roberts-Stirling~\cite{mrsa} with the factorization scale set
equal to the parton center of mass energy, $\sqrt{\hat s}$. 

To simulate the finite acceptance of detectors, we impose cuts on
observable particles in the final state. Unless otherwise stated, we
require: 
\begin{eqnarray}
\label{EQ:CUTS}\nonumber
p_T^\gamma &>& 10\ {\rm GeV}, \quad |y_\gamma| <2.5, \quad
\Delta R_{\gamma\gamma} > 0.3 \quad {\rm for \ photons}, \\
p_T^e &>& 15 \ {\rm GeV}, \quad |y_e| <2.5, \quad
\Delta R_{e\gamma} > 0.7 \quad {\rm for \ charged \ leptons},
\end{eqnarray}
and
\begin{equation}\label{EQ:PTMISS}
p\llap/_T>15~{\rm GeV}.
\end{equation}
Here, $p_T$ is the transverse momentum and $y$ the rapidity of a particle, 
and $p\llap/_T$ denotes the missing transverse momentum of the event,
defined by the imbalance to $p_T^e$ and $p_T^\gamma$ in our calculation. 
For massless particles, 
the rapidity and the pseudorapidity, $\eta$, coincide. 
\begin{equation}
\Delta R=\sqrt{(\Delta\phi)^2+(\Delta\eta)^2}
\end{equation}
denotes the separation in the pseudorapidity--azimuthal angle plane.
Without finite $p_T^\gamma$ and $\Delta R_{e\gamma}$ cuts, the cross
section for $e\nu\gamma\gamma$ production would diverge, due to the
various collinear and infrared singularities present. 

As mentioned in the Introduction, it is instructive to compare the
results obtained for $q\bar q'\to e^-\nu\gamma\gamma$ with those for
the neutral channel, $q\bar q\to e^+e^-\gamma\gamma$. In this case,
we also impose a 
\begin{eqnarray}
\label{EQ:MZCUT}
M(e^+ e^-) > 20  \ {\rm GeV}
\end{eqnarray}
cut to avoid the mass singularity from timelike
virtual photon exchange graphs. The $q\bar q\to e^+e^-\gamma\gamma$
helicity amplitudes were calculated using the same technique which we 
employed in the $e\nu\gamma\gamma$ case.
The transverse momentum and rapidity cuts listed above approximate the
phase-space region which will be covered by the upgraded 
CDF~\cite{CDFupgr} and D\O\ detectors~\cite{D0upgr}. 

Uncertainties in the energy measurement of electrons and photons are, 
unless stated otherwise, taken into account in our numerical 
simulations by Gaussian smearing with
\begin{equation}
{\sigma\over E}={0.2\over\sqrt{E}}\oplus 0.01\; ,
\label{EQ:RESOL}
\end{equation}
where the two terms are added in quadrature and $E$ is in units
of GeV. The only visible effect of the finite
energy resolution in the figures presented below arises in regions of
phase space where the cross section changes very rapidly, {\it e.g.}
around the $W$ or $Z$ boson peaks.

For the cuts listed in Eq.~(\ref{EQ:CUTS}), backgrounds to
$e\nu\gamma\gamma$ and $e^+e^-\gamma\gamma$ production are small, provided 
the two photons are well isolated from any hadronic
energy in the event. The isolation cut essentially
eliminates the backgrounds from $W\gamma+1$~jet and $W+2$~jet production
where one or both jets fragment into a photon~\cite{jim}. For $p_T^\gamma>
10$~GeV, the probability that a jet fakes a photon, $P_{j/\gamma}$, is
$10^{-3}$ or less~\cite{WGEXP}. Backgrounds from $W\gamma+$~jets and
$W+$~jets production, where one or two jets fake a photon, are then
small. The photon-photon
separation cut of $\Delta R_{\gamma\gamma}>0.3$, combined with a substantial
$p_T^\gamma$, requires a sizable invariant mass of the two-photon system and
thereby eliminates backgrounds from $\pi^0\to\gamma\gamma$ decays which might 
originate from $W/Z+1$~jet production with a leading $\pi^0$.

The geometrical acceptance of the upgraded CDF and D\O\ detectors for
muons will be similar to that for electrons. Requiring the charged lepton
to be well separated from the photons, the cross sections for 
$e\nu\gamma\gamma$ and 
$\mu\nu\gamma\gamma$ production are then nearly identical. The results
derived in the following for the electron channel therefore 
also apply to the $\mu\nu\gamma\gamma$ final state. 

\subsection{The $e^-\bar\nu\gamma\gamma$ and $W^-(\to e^-\bar\nu)
\gamma\gamma$ Cross Sections}
    
\begin{figure}
\begin{center}
\epsfxsize=4.25in
\epsfysize=4.25in
\epsfxsize=4.25in\hspace{0in}\epsffile[47 218 418 581]{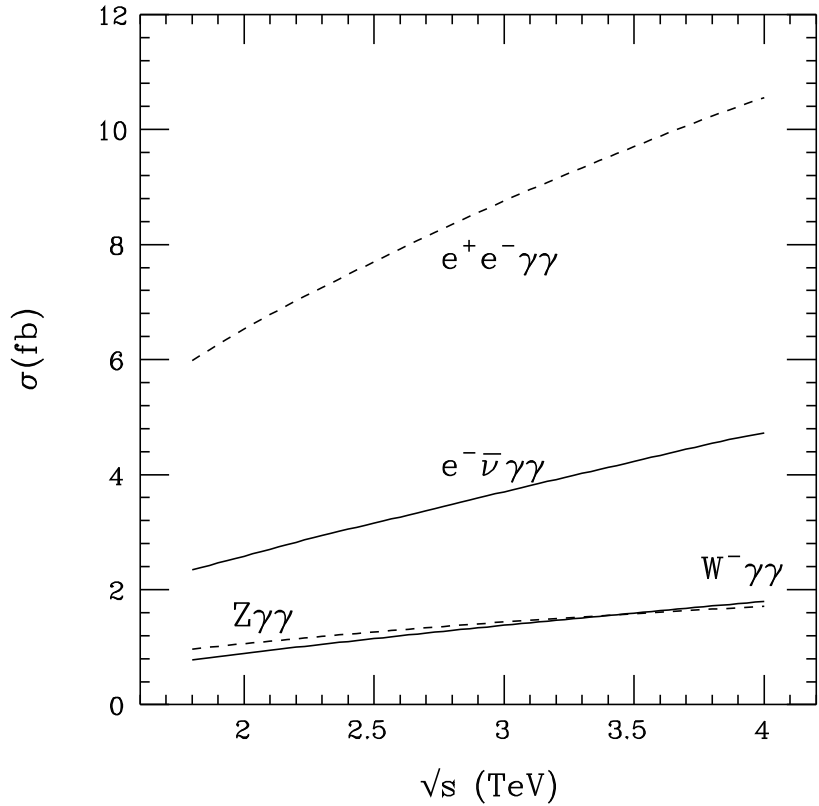}
\vspace*{0.2in}
\caption{
The total cross sections for $p\bar p \to e^- \bar\nu\gamma\gamma$ 
and $p\bar p\to W^-(\to e^-\bar\nu)\gamma\gamma$ (solid lines) as a 
function of the $p\bar p$ center of mass energy, $\protect\sqrt{s}$. 
For comparison, we also show the $p\bar p\to e^+e^-\gamma\gamma$ and
$p\bar p\to Z(\to e^+e^-) \gamma\gamma$ cross sections (dashed lines). The
acceptance cuts are summarized in Sec.~IIIA. No energy smearing is
imposed.}
\label{FIG:TOTAL}
\end{center}
\end{figure}

In Fig.~\ref{FIG:TOTAL}, 
we present the total cross sections, within the cuts of Sec.~IIIA,
for $p\bar p \rightarrow e^- \bar\nu \gamma \gamma$ and on-shell
$W^-(\to e^-\bar\nu) \gamma \gamma$ production (solid) as a function of 
the $p\bar p$ center of mass energy. For comparison, we also show the
cross sections for $e^+ e^- \gamma\gamma$ and $Z(\to e^+e^-)\gamma\gamma$ 
production (dashed). Here the on-shell $W^-(\to e^-\bar\nu)
\gamma\gamma$ and $Z(\to e^+e^-)\gamma\gamma$ cross sections have been 
calculated in the narrow $W/Z$ width approximations. 
The large differences between the on-shell and the full cross sections
arise from diagrams where one or both photons are emitted by a final 
state charged lepton. 
For $e^+e^-\gamma\gamma$ events there are also sizable contributions from
$\gamma^*\to e^+e^-$.
Contributions from these diagrams increase the cross section
by about a factor 3 (6) in the $W\gamma\gamma$ ($Z\gamma\gamma$) case
for the cuts chosen. No energy smearing effects are taken into account
in Fig.~\ref{FIG:TOTAL}. 

The rates for $p\bar p\to W^-(\to e^-\bar\nu) 
\gamma\gamma$ and $p\bar p\to Z(\to e^+e^-)\gamma\gamma$ \cite{hansobey}
are almost identical over the entire center of mass range studied
for the cuts chosen. This should be contrasted with the cross section 
ratio of $W^-(\to e^-\bar\nu)+2$~jet to $Z(\to e^+e^-)+2$~jet production 
which is about~4.6~\cite{Vjets}. The relative suppression of the
$W\gamma\gamma$ cross section can be traced to the radiation zero
which is present in $W\gamma\gamma$, but not in $Z\gamma\gamma$
production. Similarly, the $W\gamma$ cross section is 
suppressed relative to the $Z\gamma$ production rate because of the
radiation zero in $q\bar q'\to W\gamma$~\cite{BEO}. 

\begin{figure}
\begin{center}
\epsfxsize=5.0in
\epsfysize=3.5in
\epsfxsize=4.65in\hspace{0in}\epsffile[32 222 551 534]{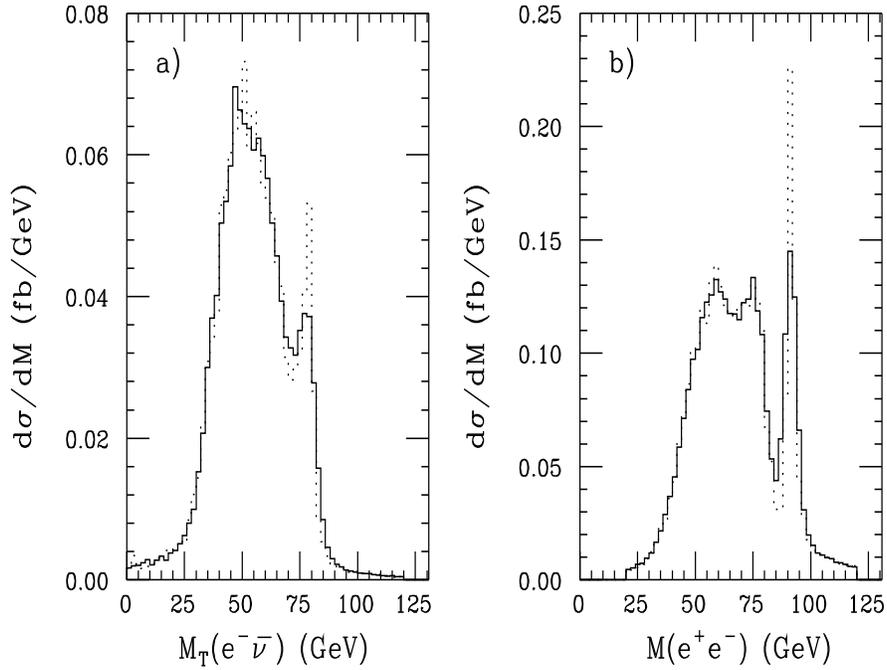}
\vspace*{0.2in}
\caption{(a) The $\protect {e\nu}$ transverse mass distribution for 
$\protect {p\bar p\to e^-\bar\nu\gamma\gamma}$, and (b) the 
$\protect {e^+e^-}$ invariant mass distribution for 
$\protect {p\bar p\to e^+e^-\gamma\gamma}$, at $\protect{\sqrt{s}=2}$~TeV. 
The cuts imposed are summarized in Sec.~IIIA. The solid and dotted 
curves give the results with and without taking into account the finite 
energy resolution of detectors (see Eq.~(\protect{\ref{EQ:RESOL}})).} 
\label{FIG:MT}
\end{center}
\end{figure}

Figure~\ref{FIG:TOTAL} shows that, although we require the charged lepton
to be well separated from the photons, radiation off the final state
charged lepton completely dominates. In order to search for a possible
radiation zero in $W\gamma\gamma$ production, it is necessary to
suppress final state bremsstrahlung more efficiently. To 
isolate the $W(\to e\nu)\gamma\gamma$ component in $p\bar p\to
e\nu\gamma\gamma$, it is useful to study the
transverse mass distribution of the $e\nu$ system which is shown in 
Fig.~\ref{FIG:MT}(a). $W(\to e\nu)\gamma\gamma$ events produce a 
$M_T(e\nu)$ distribution
which is sharply peaked at $M_T(e\nu)=m_W$. However, finite energy 
resolution effects significantly dilute this peak (see solid curve). 
On the other hand, if one or
both photons are emitted by the charged lepton, the $e\nu$ transverse
mass tends to be considerably smaller than the $W$ mass. Requiring 
\begin{equation}
M_T(e^-\nu) > 70~{\rm GeV,}
\label{EQ:TRM}
\end{equation}
eliminates most of the contributions from final state radiation. With this 
additional cut, the total and differential cross sections 
for $e\nu\gamma\gamma$ and $W(\to e\nu)\gamma\gamma$ production are almost 
identical. 

Similarly, a cut on the di-lepton invariant mass can be used to
suppress photon radiation from the final state leptons in $p\bar p\to
e^+e^-\gamma\gamma$. The $e^+e^-$ invariant mass distribution, for the
cuts summarized in Sec.~IIIA, is shown in Fig.~\ref{FIG:MT}(b). 
The two broad peaks below the $Z$ resonance region 
correspond to contributions from $Z \rightarrow e^+e^- \gamma$ 
and $Z \rightarrow e^+e^- \gamma \gamma$. Details of the structure depend
on the choices of $p_T^\gamma$ and $\Delta R_{e\gamma}$ cuts. For
\begin{equation}
M(e^+e^-) > 85~{\rm GeV,} 
\label{EQ:MZee}
\end{equation}
contributions from final state bremsstrahlung are reduced by about a 
factor of~4 for $p_T^\gamma>10$~GeV. 
Nevertheless, contributions from final state bremsstrahlung
and $\gamma^*\to e^+e^-$  are still sizeable in this case. 

\begin{figure}
\begin{center}
\epsfxsize=4.25in
\epsfysize=4.25in
\epsfxsize=4.25in\hspace{0in}\epsffile[47 211 423 574]{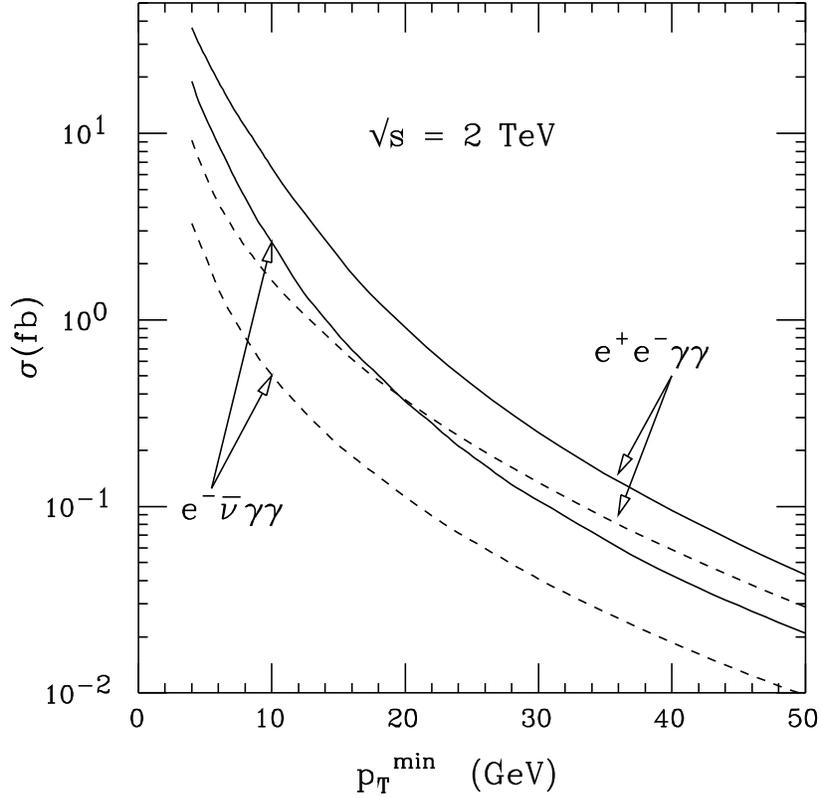}
\vspace*{0.2in}
\caption{The total cross sections for $p\bar p \rightarrow 
e^-\bar\nu\gamma\gamma$ and $p\bar p\to e^+ e^- \gamma\gamma$ 
at $\protect\sqrt s=2$~TeV as a function of the minimal photon transverse
momentum $p_T^{\rm min}$. The solid lines are for the cuts summarized in
Sec.~IIIA. For the dashed lines an additional $M_T(e\nu)>70$~GeV or
$M(e^+e^-)>85$~GeV cut has been imposed. 
No energy smearing is included here.}
\label{FIG:PTG}
\end{center}
\end{figure}

Within the cuts of Eqs.~(\ref{EQ:CUTS}) and (\ref{EQ:PTMISS}), and with
the transverse mass cut of $M_T(\ell\nu) > 70$~GeV, the total 
$\ell^\pm\nu\gamma\gamma$ ($\ell=e,\,\mu$) cross section for $p\bar p$ 
collisions at $\sqrt{s}=2$~TeV is about 2~fb. For an integrated 
luminosity of 30~fb$^{-1}$, one thus expects about 
60~$\ell^\pm\nu\gamma\gamma$ events. The cross
section depends quite sensitively on the minimum photon transverse
momentum, $p_T^{\rm min}$, however. This dependence, with and
without the transverse mass cut of Eq.~(\ref{EQ:TRM}), is shown in 
Fig.~\ref{FIG:PTG} for the $e^-\bar\nu\gamma\gamma$ cross section.
For completeness, curves for $p\bar p\to e^+e^-\gamma\gamma$ are included
as well. No energy smearing 
effects are taken into account here. Reducing the photon
transverse momentum threshold from 10~GeV to 4~GeV, the
$e^-\bar\nu\gamma\gamma$ rate, regardless of the transverse mass cut
of Eq.~(\ref{EQ:TRM}), increases by about a factor of~6. 
Due to the limited number of $e\nu\gamma\gamma$ events even at the 
highest Tevatron luminosities, the $p_T$ threshold for at least one of 
the photons should be lowered as far as 
possible in a search for the radiation zero in $W\gamma\gamma$
production. Nevertheless, in our further analysis, we shall retain
the more stringent photon transverse momentum requirement of 
$p_T^\gamma>10$~GeV for both photons (see 
Eq.~(\ref{EQ:CUTS})). As mentioned in Sec.~IIIA, 
backgrounds from $W\gamma+$~jets and $W+$~jets production, where one or 
two jets fake a photon, are then small. Furthermore, we shall impose the
mass cuts of Eq.~(\ref{EQ:TRM}) and (\ref{EQ:MZee}) unless stated
otherwise.

\subsection{Searching for the Radiation Zero} 

The general theorem of Ref.~\cite{theorem} states that, in the SM, the 
amplitude for the process $d\bar u\to W^-\gamma\gamma$ vanishes for 
\begin{eqnarray}
\label{eq:raz1}
\cos \theta_W^* = \cos \theta_{0W^-}^* = \frac {Q_d + Q_u}
{Q_d-Q_u}=-{1\over 3},
\label{EQ:ZERO}
\end{eqnarray}
\begin{figure}
\begin{center}
\epsfxsize=5.0in
\epsfysize=3.5in
\epsfxsize=5.in\hspace{0in}\epsffile[35 220 551 530]{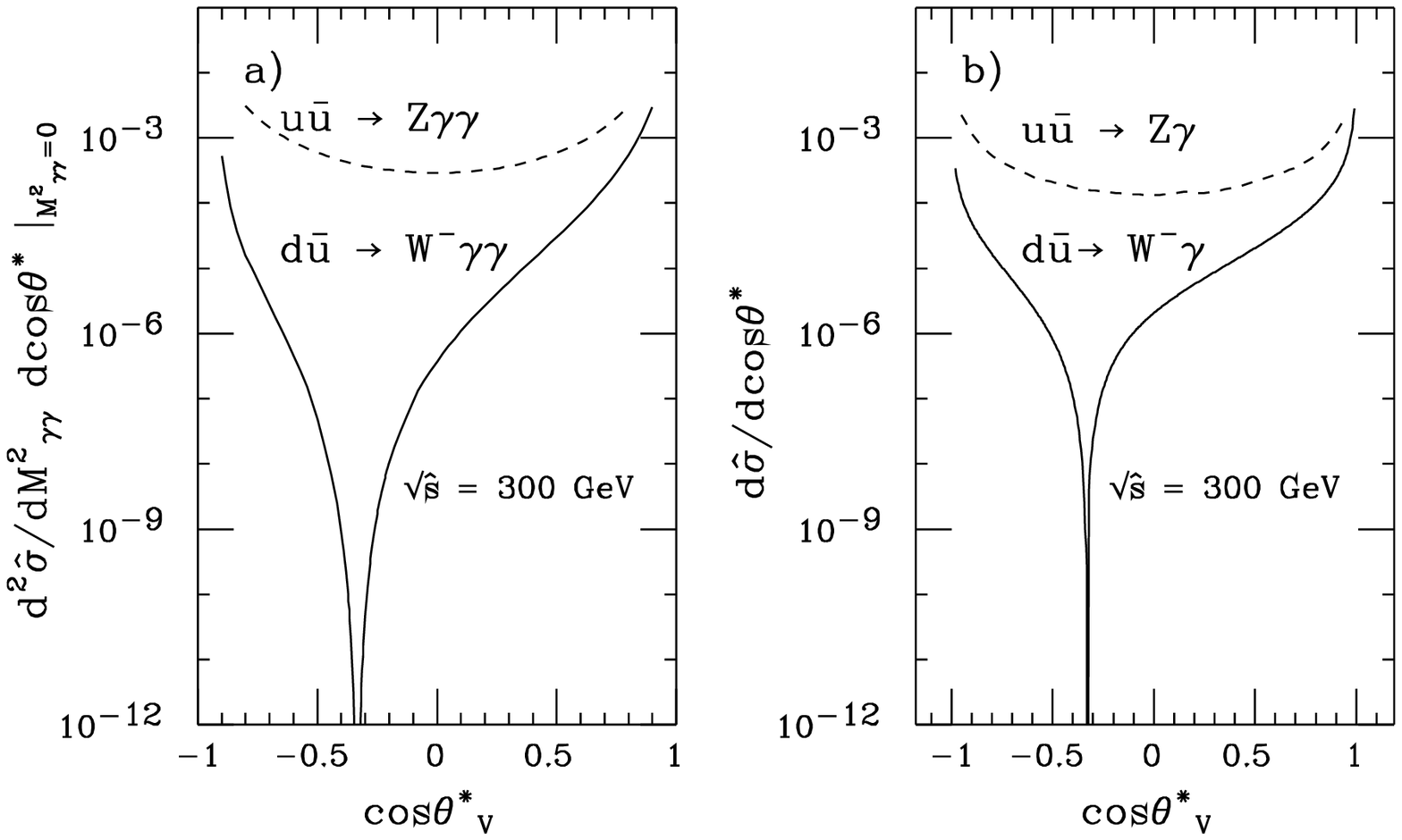}
\vspace*{0.2in}
\caption{
The angular distribution of the vector boson for the partonic
processes (a) $d \bar u \rightarrow W^- \gamma \gamma$ with
$M_{\gamma\gamma}=0$, and (b) $d \bar u \rightarrow W^- \gamma$, at 
$\protect\sqrt{\hat s}=300$~GeV (solid lines). Corresponding curves for
$u\bar u\to Z\gamma\gamma$ and $u\bar u\to Z\gamma$ (dashed lines) are 
also shown for comparison. The photon energies in the $W\gamma\gamma$
and $Z\gamma\gamma$ case are chosen to be equal. The differential cross 
sections are in arbitrary units. No cuts and no energy smearing are imposed,
and the $W$ and $Z$ bosons are treated as stable particles.}
\label{FIG:RAZ0}
\end{center}
\end{figure}
when the two photons are collinear. Here $\theta_W^*$ is the angle between 
the incoming $d$-quark and the $W$ boson, and the asterisk on a quantity
denotes that it is to be taken in the parton center of mass frame. For
$W^+\gamma\gamma$ production, the role of the $u$- and $d$-quarks in 
Eq.~(\ref{EQ:ZERO}) are interchanged, {\it i.e.} $\cos \theta_{0W^+}^*
= -\cos \theta_{0W^-}^*$. The existence of the radiation zero can be
readily verified numerically. Figure~\ref{FIG:RAZ0}(a) shows the
$\cos\theta_W^*$ distribution for the parton level process $d\bar u\to
W^-\gamma\gamma$, at a parton center of mass energy of $\sqrt{\hat 
s}=300$~GeV and for $M_{\gamma\gamma}=0$, which forces the two photons
to be collinear. In addition, the photon energies are chosen to be
equal. For unequal photon energies, qualitatively very similar results
are obtained. The vanishing of the differential cross section at
$\cos \theta_{0W^-}^* =-1/3$ is apparent. For comparison, we have also
included the $\cos\theta_Z^*$ distribution for $u\bar u\to
Z\gamma\gamma$ in Fig.~\ref{FIG:RAZ0}(a) (dashed line), and show the
$\cos\theta_W^*$ and $\cos\theta_Z^*$ distributions for $d\bar u\to
W^-\gamma$ and $u\bar u\to Z\gamma$ at $\sqrt{\hat s}=300$~GeV in 
Fig.~\ref{FIG:RAZ0}(b). The $\cos\theta_W^*$ region where the cross 
section is substantially reduced due to the zero is seen to be 
considerably larger in the $W\gamma\gamma$ case. No cuts and no energy
smearing have been imposed in Fig.~\ref{FIG:RAZ0}, and the $W$ and $Z$ bosons 
are treated as stable particles. The overall normalization of the
cross sections in each part of the figure is arbitrary. Similar results
are obtained for different parton center of mass energies.

\begin{figure}
\begin{center}
\epsfxsize=4.75in
\epsfysize=4.75in
\epsfxsize=4.75in\hspace{0in}\epsffile{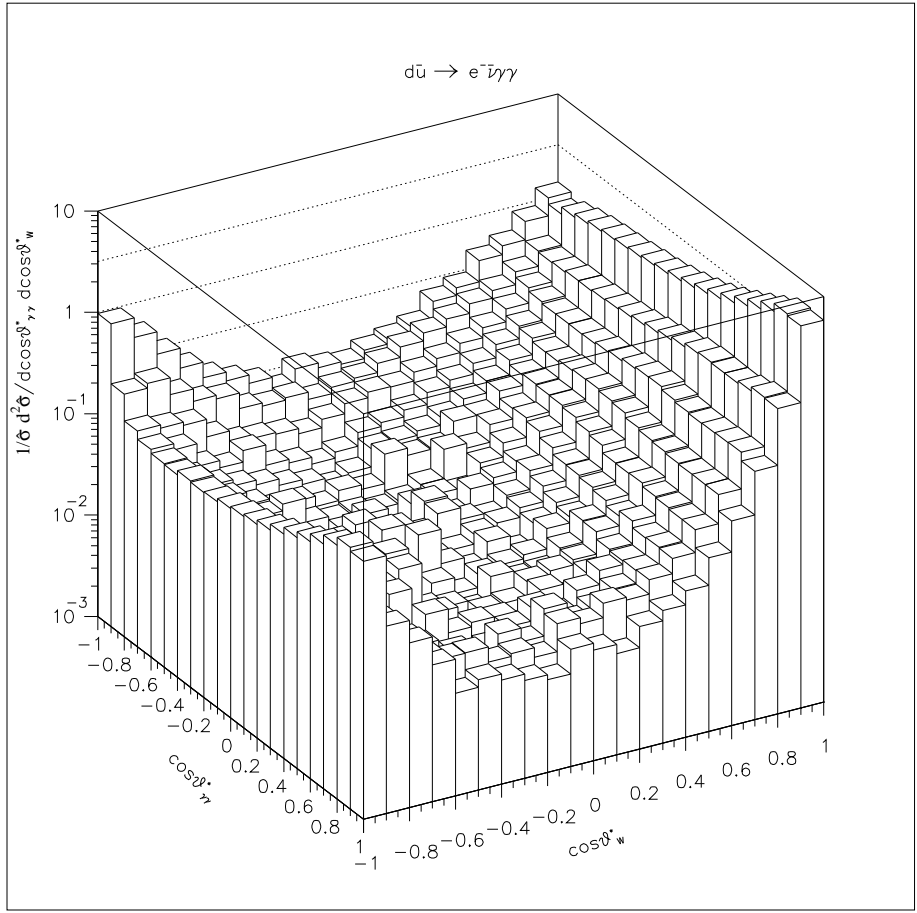}
\vspace*{0.5in}
\caption{
The normalized double differential distribution $(1/\hat\sigma)\,
(d^2\hat\sigma/d\cos\theta^*_{\gamma\gamma}\,d\cos\theta^*_W)$ for the
partonic process $d\bar u\to e^-\bar\nu\gamma\gamma$ at
$\protect{\sqrt{\hat s}=300}$~GeV. The cuts summarized in 
Eqs.~(\protect{\ref{EQ:CUTS}}), (\protect{\ref{EQ:PTMISS}})
and~(\protect{\ref{EQ:TRM}}) are imposed.}
\label{FIG:LEGO1}
\end{center}
\end{figure}

The impressive $W\gamma\gamma$ radiation zero in Fig.~\ref{FIG:RAZ0}
is washed out by the small contamination of 
$W(\to e^-\bar\nu\gamma)\gamma$ and $W\to e^-\bar\nu\gamma\gamma$
events which pass the $M_T(e^-\bar\nu)$ cut of
Eq.~(\ref{EQ:TRM}) when $W$ decays and finite $W$ width effects are 
taken into account. Binning effects reduce the radiation zero to a mere
dip as well. This is shown in Fig.~\ref{FIG:LEGO1} where we 
display the normalized double differential cross section 
$(1/\hat\sigma)(d^2\hat\sigma/d\cos\theta_W^*\,d\cos\theta^*_{\gamma\gamma})$ 
for the partonic process $d\bar u\to e^-\bar\nu\gamma\gamma$ with 
$\sqrt{\hat{s}}=300$~GeV. Here $\theta^*_{\gamma\gamma}$ is the angle 
between the two photons in the parton center of mass frame. In this figure, 
the full set of contributing Feynman diagrams, including the corrections to 
the $W$ propagator and the $WW\gamma$ and $WW\gamma\gamma$ vertices
described in Sec.~II, have been taken into account, together with the
cuts summarized in Eqs.~(\ref{EQ:CUTS}), (\ref{EQ:PTMISS}) 
and~(\ref{EQ:TRM}) which we subsequently impose in all figures.

Figure~\ref{FIG:LEGO1} demonstrates that the dip in the
$e^-\bar\nu\gamma\gamma$ differential cross section at 
$\cos\theta^*_W=-1/3$, which signals the presence of the $W\gamma\gamma$ 
radiation zero, is quite pronounced for the cuts we have chosen. It also
shows that it only gradually vanishes for increasing values of
$\theta^*_{\gamma\gamma}$. Requiring two photons with 
$\Delta R_{\gamma\gamma}>0.3$ therefore has no significant effect on 
the observability of the $W\gamma\gamma$ radiation zero. 

The significance of the dip, which signals the amplitude zero, is
potentially further
reduced by the convolution with parton distribution functions and by
the twofold ambiguity in reconstructing the parton center of mass frame.
This twofold ambiguity originates from the non-observation of the 
neutrino arising from $W$ decay. Setting the $e\nu$ invariant mass equal 
to $m_W$ leaves two solutions for the reconstructed center of mass, which
can be ordered 
according to whether the rapidity of the neutrino is larger (``plus'' 
solution) or smaller (``minus'' solution) than the rapidity of the 
electron~\cite{CHH}. 
Since the photons couple more strongly to the incoming up-type anti-quark, 
the $W^-$ boson tends to be emitted in the proton direction.
Within the SM, and as in $W\gamma$ production, the 
dominant helicity of the $W^-$ boson in $W\gamma\gamma$ production 
is $\lambda_W=-1$, 
implying that the electron is more likely to be emitted in the direction
of the parent $W$. The rapidity of the electron thus is typically larger
than that of the neutrino, and the ``minus'' solution better preserves the
dip caused by the radiation zero. In $W^+\gamma\gamma$ production, the
$W$ boson is dominantly emitted into the $\bar p$ direction, and,
consequently, the ``plus'' solution shows more similarity with the
true reconstructed parton center of mass.

\begin{figure}
\begin{center}
\epsfxsize=4.75in
\epsfysize=4.75in
\epsfxsize=4.75in\hspace{0in}\epsffile{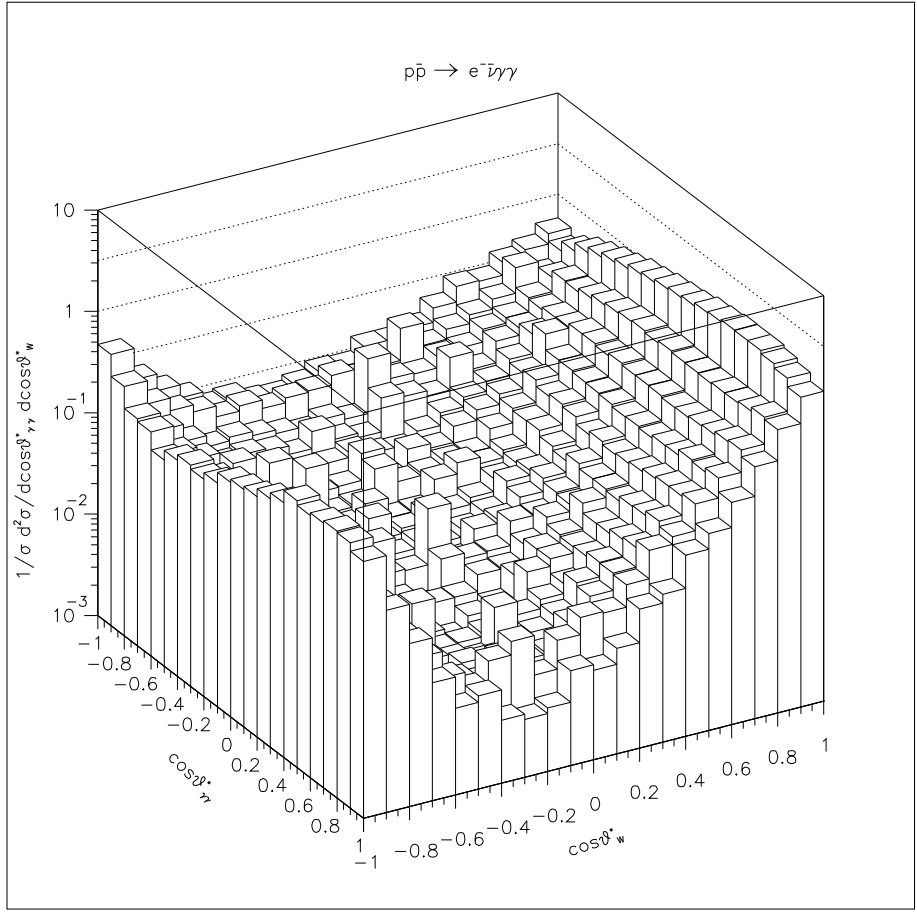}
\vspace*{0.5in}
\caption{
The normalized double differential distribution $(1/\sigma)\,
(d^2\sigma/d\cos\theta^*_{\gamma\gamma}\,d\cos\theta^*_W)$ for the
process $p\bar p\to e^-\bar\nu\gamma\gamma$ at
$\protect{\sqrt{s}=2}$~TeV, using the ``minus'' solution for the
reconstructed parton center of mass. The cuts summarized in 
Eqs.~(\protect{\ref{EQ:CUTS}}), (\protect{\ref{EQ:PTMISS}})
and~(\protect{\ref{EQ:TRM}}) are imposed.}
\label{FIG:LEGO2}
\end{center}
\end{figure}

The normalized double differential distribution $(1/\sigma)\,
(d^2\sigma/d\cos\theta^*_{\gamma\gamma}\,d\cos\theta^*_W)$ for the process
$p\bar p\to e^-\bar\nu\gamma\gamma$ at $\sqrt{s}=2$~TeV is shown
in Fig.~\ref{FIG:LEGO2}, 
using the ``minus'' solution for the reconstructed parton 
center of mass. The distribution is seen to be quite similar to
the corresponding partonic differential cross section shown in 
Fig.~\ref{FIG:LEGO1}. The convolution with the parton distribution
functions therefore has only a minor effect on the observability of the
radiation zero. Likewise, the reconstruction of the parton center of
mass frame does not affect the significance of the dip much, provided
that the appropriate solution for the longitudinal momentum of the
neutrino is used, and the missing transverse momentum is well measured
(see below). 

\begin{figure}
\begin{center}
\epsfxsize=5.0in
\epsfysize=3.5in
\epsfxsize=5.0in\hspace{0in}\epsffile[38 218 551 530]{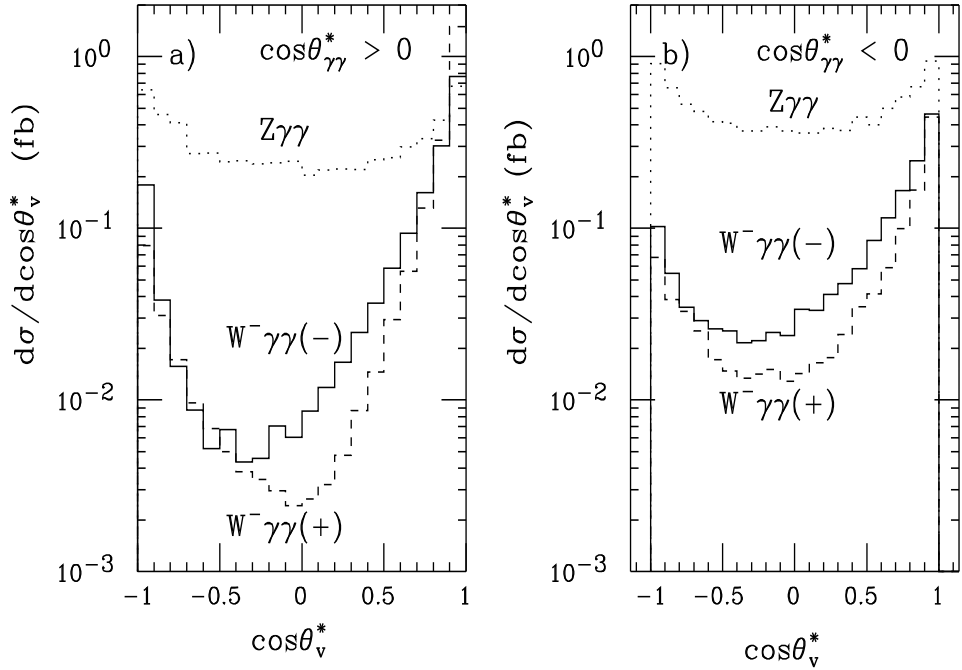}
\vspace*{0.2in}
\caption{
The $\cos \theta^*_W$ distribution for $p\bar p\to 
e^-\bar\nu\gamma\gamma$ at $\protect{\sqrt{s}=2}$~TeV, for (a) 
$\cos\theta^*_{\gamma\gamma}>0$ and (b) $\cos\theta^*_{\gamma\gamma}<0$.
The solid (dashed) line is for the ``minus'' (``plus'')
solution of the reconstructed parton center of mass frame.
The dotted line displays the $\cos\theta_Z^*$ distribution for $p\bar
p\to e^+ e^- \gamma \gamma$ for comparison. The cuts summarized in 
Eqs.~(\protect{\ref{EQ:CUTS}}), (\protect{\ref{EQ:PTMISS}})
and~(\protect{\ref{EQ:TRM}}) are imposed.
In the $Z\gamma\gamma$ case, the cuts listed in Eqs.~(\protect{\ref{EQ:CUTS}})
and~(\protect{\ref{EQ:MZee}}) are applied.}
\label{FIG:ERAZ0}
\end{center}
\end{figure}

For the limited number of $e\nu\gamma\gamma$ events expected in future
Tevatron runs, it will be impossible to map out the double differential
distribution shown in Fig.~\ref{FIG:LEGO2}. However, since the dip
signaling the radiation zero disappears only gradually with increasing
values of $\theta^*_{\gamma\gamma}$, most of the information present in
$d^2\sigma/d\cos\theta^*_{\gamma\gamma}\,d\cos\theta^*_W$ is contained in
the $\cos\theta^*_W$ distributions for events with 
$\cos\theta^*_{\gamma\gamma}>0$ versus $\cos\theta^*_{\gamma\gamma}<0$.
These two $\cos\theta^*_W$ distributions are shown in Fig.~10, for both the 
``plus'' and the ``minus'' solution of the reconstructed parton center 
of mass frame.  For comparison, Fig.~\ref{FIG:ERAZ0} also shows 
the $\cos\theta^*_Z$ distribution for $p\bar p\to e^+e^-\gamma\gamma$ 
with $M(e^+e^-)>85$~GeV and the cuts of Eq.~(\ref{EQ:CUTS}). For 
$\cos\theta^*_{\gamma\gamma}>0$ and using the ``minus'' solution, the 
$\cos\theta^*_W$ distribution displays a pronounced dip located at 
$\cos\theta^*_W\approx -1/3$. For the ``plus'' solution the minimum is
shifted to $\cos\theta^*_W\approx 0$. In contrast, requiring
$\cos\theta^*_{\gamma\gamma}<0$, the dip is drastically reduced, and the 
differential
cross section at $\cos\theta^*_W\approx -1/3$ is about one order of
magnitude larger than for $\cos\theta^*_{\gamma\gamma}>0$ 
(see Fig.~\ref{FIG:ERAZ0}(b)). 
The large difference in the $\cos\theta^*_W$ distribution for 
$\cos\theta^*_{\gamma\gamma}>0$  and $\cos\theta^*_{\gamma\gamma}<0$
becomes more apparent by comparing the $\cos\theta^*_Z$ distribution in
$Z\gamma\gamma$ production in the two regions. Unlike the situation
encountered in $W\gamma\gamma$ production, the $\cos\theta^*_Z$
distributions for $\cos\theta^*_{\gamma\gamma}>0$  and 
$\cos\theta^*_{\gamma\gamma}<0$ are very similar. 

Determining the $\cos\theta^*_W$ distribution requires measurement
of the transverse momentum of the neutrino produced in the $W$ decay. 
In $e\nu\gamma\gamma$ production, the 
neutrino transverse momentum is identified with the missing transverse 
energy, $E\llap/_T$, in the event. In future Tevatron runs, one expects 
up to ten interactions per bunch crossing~\cite{marriner}. Multiple 
interactions per crossing significantly worsen the $E\llap/_T$ 
resolution, and thus tend to wash out the dip signaling the 
radiation zero. We have not included missing transverse energy 
resolution effects in our simulations, since the number of interactions per 
crossing, and hence the $E\llap/_T$ resolution, sensitively depend on 
the future Tevatron accelerator parameters which are difficult to 
foresee at present. 

Due to the negative impact of multiple interactions on the missing
transverse energy resolution, it is advantageous to search for a
kinematic variable which exhibits a clear signal of the radiation zero
but does not depend on the neutrino momentum. The $y_{\gamma\gamma}-y_e$
distribution is a possible candidate for such a variable. Here $y_e$ is 
the electron rapidity and $y_{\gamma\gamma}$ denotes the rapidity of the 
two-photon system in the laboratory frame.

In Ref.~\cite{rapraz} it was found that photon lepton rapidity
correlations are a useful tool to search for the radiation zero in
$W\gamma$ production. The distribution of the rapidity difference,
$\Delta y(\gamma,e)=y_\gamma-y_e$, which does not require
knowledge of the missing transverse energy or the longitudinal momentum 
of the neutrino, clearly displays the SM radiation zero in form of a
dip. In the parton center of mass frame, the photon and $W$ boson in
$q\bar q'\to W^-\gamma$ are back to back. Due to the radiation zero,
the photon and $W$ rapidity
distributions in the parton center of mass frame, $d\sigma/dy^*_\gamma$ 
and $d\sigma/dy^*_W$, display pronounced dips located at 
\begin{eqnarray}
y^*_\gamma&  = & {1\over 2}\ln 2\approx 0.35, \\
y^*_W     &\approx& -0.05.
\end{eqnarray}
If $W$ mass effects could be ignored, one would expect that 
$y^*_\gamma = -y^*_W$.
Since differences of rapidities are invariant under longitudinal boosts,
the difference of the photon and the $W$ rapidity in the laboratory frame
then exhibits a dip at
\begin{equation}
\Delta y(\gamma,W)=y_\gamma-y_W=y^*_\gamma-y^*_W\approx 0.4.
\end{equation}
As discussed earlier, the dominant $W$ helicity in $W^\pm\gamma$ production 
is $\lambda_W=\pm 1$, implying that the charged lepton tends to be emitted
in the direction of the parent $W$, and thus reflects most of its
kinematic properties. The dip signaling the presence of the radiation
zero therefore manifests itself in the $\Delta y(\gamma,\ell)$
distribution. Since the average rapidity of the lepton and the $W$ are
slightly different, the location of the minimum is shifted to 
$\Delta y(\gamma,\ell)\approx 0.1$.

\begin{figure}
\begin{center}
\epsfxsize=4.0in
\epsfysize=4.0in
\epsfxsize=4.75in\hspace{0in}\epsffile{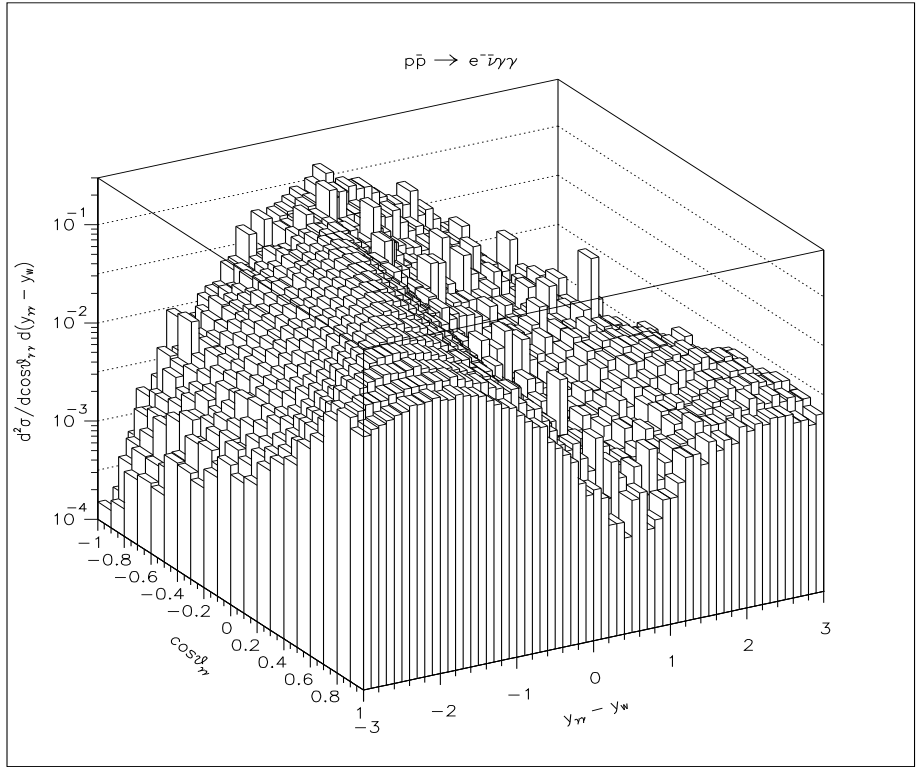}
\vspace*{0.5in}
\caption{The double differential distribution
$d^2\sigma/d\cos\theta_{\gamma\gamma}\,d(y_{\gamma\gamma}-y_W)$ for the
process $p\bar p\to e^-\bar\nu\gamma\gamma$ at
$\protect{\sqrt{s}=2}$~TeV, using the ``minus'' solution for the
longitudinal neutrino momentum. The cuts summarized in 
Eqs.~(\protect{\ref{EQ:CUTS}}), (\protect{\ref{EQ:PTMISS}})
and~(\protect{\ref{EQ:TRM}}) are imposed.}
\label{FIG:LEGO3}
\end{center}
\end{figure}

The radiation zero in $W\gamma\gamma$ production occurs at exactly the 
same rapidity as the zero in $W\gamma$ production, when the photons are 
collinear. One therefore expects that the $\Delta y(\gamma\gamma,W)$ 
distribution displays a clear dip for photons with a small opening angle, 
$\theta_{\gamma\gamma}$, in the
laboratory frame, {\it i.e.} at $\cos\theta_{\gamma\gamma}\approx 1$. In
Fig.~\ref{FIG:LEGO3} we show the double differential distribution
$d^2\sigma/d\cos\theta_{\gamma\gamma}\,d(y_{\gamma\gamma}-y_W)$, using
the ``minus'' solution for the longitudinal neutrino momentum. For 
$\Delta y(\gamma\gamma,W)\approx 0.4$, a clear dip is visible for
$\cos\theta_{\gamma\gamma}$ values close to one. The dip gradually 
vanishes for larger opening angles between the two photons, leading to 
a ``canyon'' in the double differential distribution. Due to the finite
invariant mass of the $\gamma\gamma$ system for non-zero values of
$\theta_{\gamma\gamma}$, the location of the minimum in 
$\Delta y(\gamma\gamma,W)$ varies slightly with $\cos\theta_{\gamma\gamma}$. 

\begin{figure}
\begin{center}
\epsfxsize=5.0in
\epsfysize=3.5in
\epsfxsize=5.0in\hspace{0in}\epsffile[38 216 553 530]{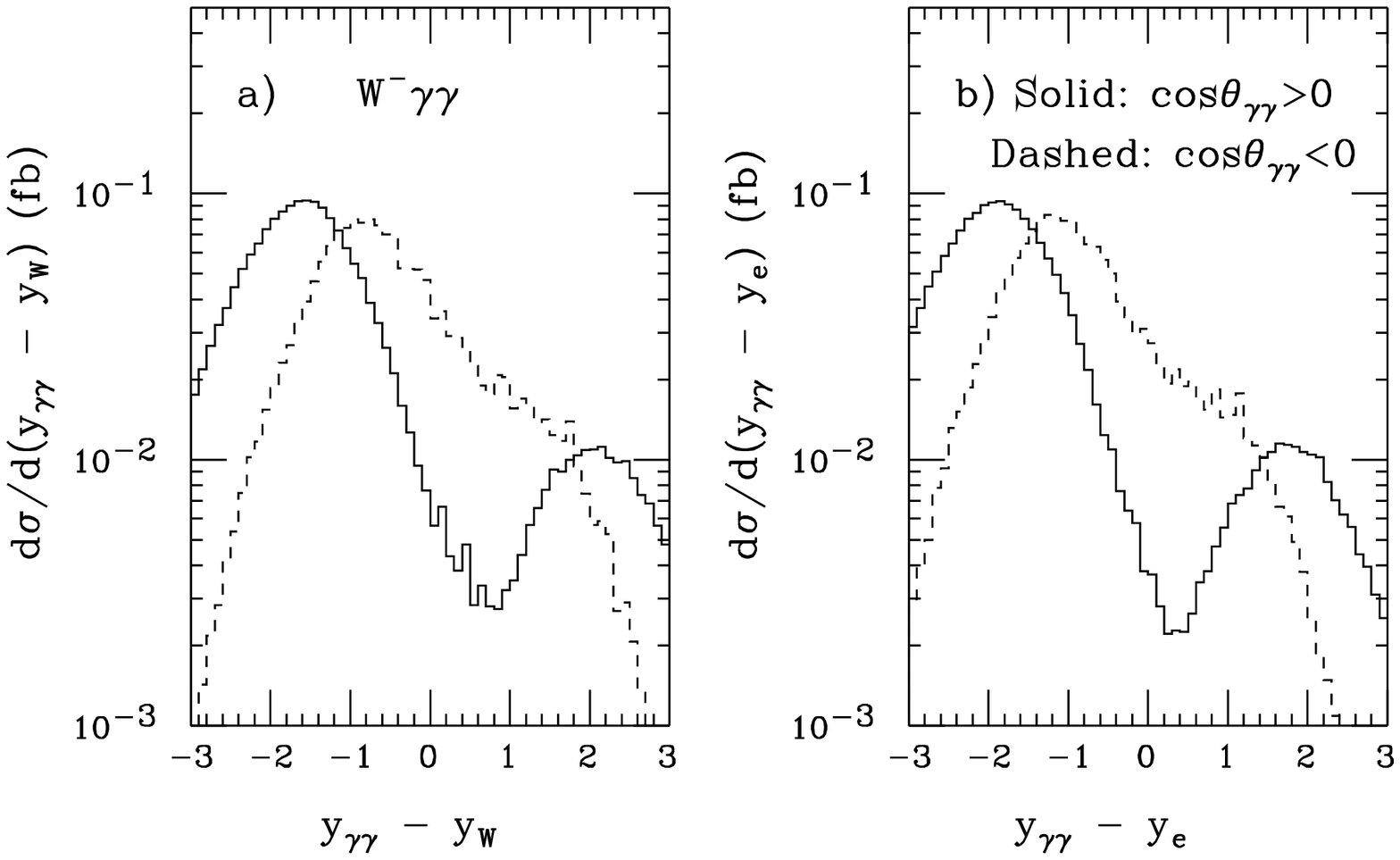}
\vspace*{0.2in}
\caption{Rapidity difference distributions for $p\bar p\to e^-\bar 
\nu\gamma\gamma$ at $\protect{\sqrt{s}=2}$~TeV. Part (a) shows the
$y_{\gamma\gamma}-y_W$ spectrum, while part (b) displays the 
$y_{\gamma\gamma}-y_e$ distribution. The solid (dashed) curves are for
$\cos\theta_{\gamma\gamma}>0$ ($\cos\theta_{\gamma\gamma}<0$). 
The cuts summarized in Eqs.~(\protect{\ref{EQ:CUTS}}), 
(\protect{\ref{EQ:PTMISS}}) and~(\protect{\ref{EQ:TRM}}) are imposed.} 
\label{FIG:XRAZ0}
\end{center}
\end{figure}

Since the dip vanishes gradually with decreasing
$\cos\theta_{\gamma\gamma}$, it is useful to consider the $\Delta 
y(\gamma\gamma,W)$ distribution for $\cos\theta_{\gamma\gamma}>0$ and
$\cos\theta_{\gamma\gamma}<0$. Figure~\ref{FIG:XRAZ0}(a) 
displays a pronounced dip in
$d\sigma/d\Delta y(\gamma\gamma,W)$ for $\cos\theta_{\gamma\gamma}>0$,
located at $\Delta y(\gamma\gamma,W)\approx 0.7$ (solid line). 
In contrast, for $\cos\theta_{\gamma\gamma}<0$, the $\Delta 
y(\gamma\gamma,W)$ distribution does not exhibit a dip (dashed line).
The $\Delta y(\gamma\gamma,W)$ distribution for
$\cos\theta_{\gamma\gamma}>0$ thus plays a role similar to the $\Delta 
y(\gamma,W)$ distribution in $W\gamma$ production. In the dip 
region, the differential cross section for $\cos\theta_{\gamma\gamma}<0$ 
is about one order of magnitude
larger than for $\cos\theta_{\gamma\gamma}>0$. In addition, the 
$\Delta y(\gamma\gamma,W)$ distribution extends to significantly higher 
$y_{\gamma\gamma}-y_W$ values if one requires 
$\cos\theta_{\gamma\gamma}>0$. This reflects the narrower rapidity 
distribution of the two-photon system for $\cos\theta_{\gamma\gamma}<0$,
due to the larger invariant mass of the system when the two photons are 
well separated.

Exactly as in the $W\gamma$ case,
the dominant helicity of the $W$ boson in $W^\pm\gamma\gamma$ production
is $\lambda_W=\pm 1$. One therefore
expects that the distribution of the rapidity difference of the
$\gamma\gamma$ system and the electron is very similar to the
$y_{\gamma\gamma}-y_W$ distribution and shows a clear signal of
the radiation zero for positive values of $\cos\theta_{\gamma\gamma}$.
The $y_{\gamma\gamma}-y_e$ distribution, shown in Fig.~\ref{FIG:XRAZ0}(b),
indeed clearly displays these features. Due to the finite difference
between the electron and the $W$ rapidities, the location of the minimum
is again slightly shifted. The $\Delta
R_{\gamma\gamma}>0.3$ cut has only little effect on the significance of the
dip.

The characteristic differences between the $\Delta y(\gamma\gamma,e)=
y_{\gamma\gamma}-y_e$ distribution for $\cos\theta_{\gamma\gamma}>0$ and
$\cos\theta_{\gamma\gamma}<0$ are also reflected in the cross section
ratio
\begin{equation}
{\cal R}= {\int_{\Delta y(\gamma\gamma,e)>-1}d\sigma\over
\int_{\Delta y(\gamma\gamma,e)<-1}d\sigma}~,
\end{equation}
which may be useful for small event samples. Many experimental
uncertainties cancel in ${\cal R}$. For $\cos\theta_{\gamma\gamma}>0$
one finds ${\cal R}\approx 0.25$, whereas for
$\cos\theta_{\gamma\gamma}<0$ ${\cal R}\approx 1.06$. 

Our calculations have all been carried out in the Born approximation.
The complete NLO
QCD corrections to $W\gamma\gamma$ production have not been
calculated yet; only the hard jet corrections to $W(\to e\nu)\gamma\gamma$
production are known~\cite{zhou}. It is reasonable, however, to take the 
known NLO QCD correction to $W\gamma$ production as a guide~\cite{NLO,NLOTWO}.
At ${\cal O}(\alpha_s)$ the virtual corrections only enter via their 
interference with the Born amplitude, and thus the radiation zero is 
preserved in the product. Among the real emission corrections, 
quark-antiquark annihilation processes dominate at Tevatron energies.
According to the theorem of Ref.~\cite{theorem}, extra gluon emission, 
{\it i.e.} the process $q\bar q'\to W^\pm n\gamma g$, exhibits a 
radiation zero at $\cos\theta_W^*=\pm 1/3$ if the gluon is collinear 
to all emitted photons, and also in the
soft gluon limit, $E_g\to 0$ (again, provided the photons are collinear).
This leaves quark-gluon initiated processes to potentially spoil the 
radiation zero. They are still suppressed at the Tevatron, however,
especially when a large photon-jet separation is required. As a result,
we expect the dip signaling the radiation amplitude zero to remain 
observable, at Tevatron energies, once NLO corrections are included.

At the LHC ($pp$ collisions at $\sqrt{s}=14$~TeV~\cite{keil}), the bulk
of the QCD corrections to $W\gamma\gamma$ production originates from
quark gluon fusion and the kinematical region where the final state
quark radiates a soft $W$ boson which is almost collinear to the quark.
Events which originate from this phase space region usually contain a
high $p_T$ jet. Since there is no radiation zero present in the
dominating $qg\to W\gamma\gamma q'$ and $g\bar q'\to W\gamma\gamma q$
processes, it is likely that QCD corrections considerably obscure the 
signal of the
$W\gamma\gamma$ radiation zero at the LHC, as in the $W\gamma$
case~\cite{rapraz}. This conjecture is supported by the large relative
cross section of $W\gamma\gamma+1$~jet production as compared to 
$W\gamma\gamma$ production reported in 
Ref.~\cite{zhou}. Although a jet veto should help reducing the size of
the QCD corrections, NLO QCD corrections to $W\gamma\gamma+0$~jet
production may still significantly reduce the observability of the
radiation zero for jet definition criteria which are realistic at LHC
energies. We therefore do not consider $e\nu\gamma\gamma$ production at 
the LHC in more detail here.

\section{Summary and Conclusions}

We have presented a calculation of the process $p\bar p\to
e\nu\gamma\gamma$ including final state bremsstrahlung diagrams and
finite $W$ width effects, and explored the prospects to observe the
radiation zero predicted by the SM for $p\bar p\to W\gamma\gamma$ in
future Tevatron collider experiments. In order to obtain a gauge
invariant scattering amplitude, the imaginary parts of the $WW\gamma$ 
triangle graphs and 
$WW\gamma\gamma$ box diagrams have to be included, in addition to
resumming the imaginary contributions to the $W$ vacuum polarization.
The imaginary parts of the triangle and box diagrams were found to change
the lowest order $WW\gamma$ and $WW\gamma\gamma$ vertex functions by a
factor $(1+i\Gamma_W/m_W)$ for the momentum configuration
relevant for the process $q\bar q'\to e\nu\gamma\gamma$. A gauge
invariant result for the $q\bar q'\to e\nu\gamma\gamma$ amplitude is 
then obtained by replacing all $W$ propagators, $WW\gamma$ and 
$WW\gamma\gamma$ vertices by the full expressions of 
Eqs.~(\ref{Wprop}), (\ref{WWgvertex}) and~(\ref{WWggvertex}),
respectively. The same prescription also ensures that the Ward
identities relating the $WWn\gamma$ and $WW(n-1)\gamma$, $n\geq3$,
vertex functions
are fulfilled, and thus yield a gauge invariant amplitude
for $q\bar q'\to e\nu + n\gamma$ with $n\geq 3$, without taking into
account one-loop corrections to these higher vertex functions.

The SM predicts the existence of a radiation zero in $q\bar q'\to
W^\pm\gamma\gamma$ at $\cos\theta^*_W=\pm 1/3$ if the two photons are
collinear. Here $\theta_W^*$ is the angle between the $W$ and the
incoming quark in the parton center of mass frame.
Since it is very difficult to experimentally separate two
collinear photons, one has to search for a signal of the radiation zero
which survives an explicit photon--photon separation requirement.
Contributions from Feynman diagrams where one or both photons are
emitted by the final state charged lepton eliminate the radiation
zero and therefore need to be suppressed by suitable cuts. We found that
a large lepton--photon separation of $\Delta R_{e\gamma}>0.7$,
together with a cut on the $e\nu$ transverse mass of $M_T(e\nu)>70$~GeV
suppresses these contributions sufficiently. 

The $W\gamma\gamma$ radiation zero is signaled by a pronounced 
dip in the $\cos\theta_W^*$ distribution if one requires
$\cos\theta^*_{\gamma\gamma}>0$. In 
contrast, no dip is present for $\cos\theta^*_{\gamma\gamma}<0$. In
order to measure the $\cos\theta^*_W$ distribution, the parton center of
mass frame has to be reconstructed. Since the neutrino originating from
the $W$ decay is not observed in the detector, this is only possible modulo
a twofold ambiguity. The two solutions can be ordered according to whether 
the reconstructed rapidity of the neutrino is
larger (``plus'' solution) or smaller (``minus'' solution) than the 
rapidity of the charged lepton. For $W^-\gamma\gamma$
($W^+\gamma\gamma$) production, the ``minus'' (``plus'') solution is
found to best represent the expected kinematical features. 

When searching for the radiation zero in $W\gamma\gamma$ production it is 
advantageous to consider alternate variables which, unlike the 
$\cos\theta_W^*$ distribution, do not depend on the neutrino momentum. 
The rapidity difference between the two-photon system and the electron,
{\it i.e.} the $y_{\gamma\gamma}-y_e$
distribution, fulfills this requirement. It was found to exhibit a 
pronounced dip which signals the presence of the radiation zero if a  
$\cos\theta_{\gamma\gamma}>0$ cut is imposed ($\theta_{\gamma\gamma}$ 
being the opening angle between the two photons in the laboratory 
system). As expected, the $y_{\gamma\gamma}-y_e$ distribution shows no 
dip for $\cos\theta_{\gamma\gamma}<0$. A photon--photon separation
cut of $\Delta R_{\gamma\gamma}>0.3$ has little effect on
the observability of the radiation zero. Although we have restricted our
discussion to $e\nu\gamma\gamma$ production, our results also apply to
$p\bar p\to\mu\nu\gamma\gamma$.

The conditions for which one expects a radiation zero in the SM $q\bar
q'\to W\gamma\gamma$ and $q\bar q'\to W\gamma$ amplitudes and the
location of the zeros are closely related: the four-momentum of the
photon in $W\gamma$ production simply has to be replaced by the
four-momentum of the $\gamma\gamma$ system in the $W\gamma\gamma$ case
with the additional requirement that the two photons are collinear. We
have demonstrated that a similar replacement in the $W\gamma$ 
photon--lepton rapidity difference distribution, with the less stringent 
requirement on the opening angle between the photons of
$\cos\theta_{\gamma\gamma}>0$, is in fact sufficient to produce an 
observable signal of the $W\gamma\gamma$ radiation zero (see 
Fig.~\ref{FIG:XRAZ0}). 

NLO QCD corrections to $p\bar p\to W\gamma\gamma$ are expected to be
modest at Tevatron energies. Given a sufficiently large integrated
luminosity, experiments at the Tevatron studying correlations between
the rapidity of the photon pair and the charged lepton therefore offer an
excellent opportunity to search for the SM radiation zero in hadronic
$W\gamma\gamma$ production. 

\acknowledgements

We would like to thank M.~Demarteau, S.~Errede and G.~Landsberg for many 
stimulating discussions. One of us (U.B.) would like to thank the
Fermilab Theory Group for its warm hospitality during various stages of
this work.  
This research was supported in part by the University of Wisconsin 
Research Committee with funds granted by the Wisconsin Alumni Research 
Foundation and by the Davis Institute for High Energy Physics, 
in part by the U.~S.~Department of Energy under Grant 
No.~DE-FG02-95ER40896 and No.~DE-FG03-91ER40674, and in part by the
National Science Foundation under Grant No. PHY9600770.

%
%

%
\end{document}